\documentclass[%
 reprint,
 amsmath,amssymb,
 aps,
pra,
]{revtex4-2}

\usepackage[T1]{fontenc}

\usepackage{graphicx}
\usepackage{dcolumn}
\usepackage{bm}
\usepackage{hyperref}
\usepackage{epstopdf}

\usepackage{multirow}

\newcommand{\dif}{\mathrm{d}}

\begin{document}

\preprint{}

\title{Empty and filled vortices in squeezed ${}^{39}$K Bose-Bose liquid drops}

\author{Ivan Poparić}
\affiliation{%
  Faculty of Science, University of Split, Ruđera Boškovića 33, HR-21000 Split, Croatia,\\
  Departament de Física, Universitat Politècnica de Catalunya, Campus Nord B4-B5, E-08034 Barcelona, Spain
}%
\author{Leandra Vranješ Markić}%
\affiliation{%
  Faculty of Science, University of Split, Ruđera Boškovića 33, HR-21000 Split, Croatia}%


\author{Jordi Boronat}
\affiliation{
  Departament de Física, Universitat Politècnica de Catalunya, Campus Nord B4-B5, E-08034 Barcelona, Spain
}%


\date{\today}

\begin{abstract}
Using density functional theory, we have theoretically studied the formation and 
the stability of vortices in quantum liquid droplets composed of a mixture of 
hyperfine states of potassium. Following the experimental setup that produced 
quantum droplets for the first time, we work with squeezed drops that are 
compressed in one direction. By squeezing the drops even more, towards a 
quasi-two dimensional geometry, we study the minimum atom number able to show a 
stable vortex and obtain that this number is significantly smaller than previous 
predictions for spherical droplets. The reduction of the critical atom number 
for forming a stable vortex could make their experimental observation in these 
droplets, which is still lacking, more feasible. Contrary to results obtained in 
heteronuclear mixtures, where the energetically preferred vortices are partially 
filled with the species not participating in the rotation, our results show a 
relevant stability island of fully empty vortices. Increasing the number of 
particles in the drop and the speed of rotation, we estimate the transition line 
between empty and filled vortices.
\end{abstract}

\maketitle

\section{\label{sec:I}Introduction}

Vortices, quantized topological excitations, are one of the clearest signatures of superfluidity. They have been observed and intensively studied in both helium droplets \cite{Fiszdon_1991,barenghi2001quantized} and trapped ultracold atoms \cite{fetter_rotating_bec_2009, Pitaevskii-Stringari,Anderson2010,Sharma2024}. Nowadays, it is also possible to trap more than one species of ultracold atoms simultaneously, which allows for the study of their much richer behavior under rotation~\cite{Saarikoski_2010}. Vortices with filled cores were predicted and observed in repulsive trapped mixtures of cold atoms \cite{williams_1999, Matthews_1999, Anderson_2000}. The cores were filled with the component which was not set in rotation. Experimentally, it was possible to remove partially or fully the component filling the core and observe the shrinking of the vortex core~\cite{Anderson_2000}. The capture of impurities in the core of quantum vortices was also used in the past to visualize vortex arrays in superfluid $^4$He~\cite{Bewley2006}.

Multicomponent systems lead to other features unobserved in single component systems. Multiply quantized vortices are normally unstable in single component superfluids and decay rapidly in singly quantized vortices. Mixtures of Bose gases confined in harmonic traps provide dynamical instabilities and splitting~\cite{kuopanportti_2019}, which are not available in single component systems. However, it has been predicted that due to interactions, the component which fills the core can slow down the splitting or even stabilize the multiply quantized  vortices~\cite{Bargi_2007,christensson_2008,Kuopanportti_2015,Patrick_2023}. These predictions are made for tightly confined systems in one direction, allowing for a two-dimensional approximation. Since in immiscible bosonic binary mixtures the vortex in one component is filled with the other one, it has been recently shown that vortices can be hydrodinamically treated as massive core vortices~\cite{Richaud_2020}. Within this approach, it has been possible to show changes in the dynamics of quantum vortices, allowing the stabilization of doubly quantized vortices in flat geometries~\cite{Richaud_2023} and exchange of mass between vortices, thus creating bosonic Josephson junctions~\cite{Bellettini_2024}.

Ten years ago, the formation and stability of ultradilute liquid droplets in  Bose-Bose mixtures, was predicted by Petrov~\cite{petrov_quantum_2015}.
Few years later, their existence was confirmed in experiments with hyperfine $^{39}$K mixtures and soon after with $^{39}$K-$^{87}$Rb, $^{41}$K-$^{87}$Rb and $^{23}$Na-$^{87}$Rb mixtures~\cite{cabrera_quantum_2018, semeghini_self-bound_2018, derrico_observation_2019, Cavicchioli_2025, Guo_2021}. These droplets arise from the interplay between the attractive interspecies interactions and quantum fluctuations, unlike classical or Helium droplets that result from the interplay between the short-range repulsive and long-range attractive component of the interatomic potential.
As their name suggests, ultradilute droplets have orders of magnitude lower density than liquid $^4$He. Furthermore, unlike liquid $^4$He, they demand a certain critical number of atoms to achieve self-binding. The lowest number of atoms that was achieved in experiments in Tarruell's group, which confined the droplet in one direction, was around 3500 \cite{cabrera_quantum_2018}, about an order of magnitude lower than in the case of spherical droplets in the Florence experiment from Ref.~\cite{ semeghini_self-bound_2018}. Some of us have recently shown that this number can be reduced by further squeezing the droplet in one direction, towards a two-dimensional (2D) state, reducing the critical number for self-binding to 1000 atoms in the case of the strongest attractive interaction between the atoms considered~\cite{sanuy_squeezing_2024}.

Although of significant interest, vortices have not been observed so far in experiments with ultradilute droplets. 
Density functional studies~\cite{ancilotto_self-bound_2018, caldara_vortices_2022} of $^{41}$K-$^{87}$Rb droplets have predicted stable vortex states for very large droplets of the order of 10$^6$ atoms, which is significantly larger than the droplets studied in experiments so far. They showed that the formation of linear vortices in the heavier species was energetically most favorable,
resulting in the formation of partially filled cores, similarly to the case of trapped repulsive Bose-Bose mixtures.

Most other theoretical studies of vortices in ultradilute droplets are performed in an effective 2D model in a trap ~\cite{examilioti_ground_2020, nikolaou_rotating_2023, nikolaou_rotating_2023-1, tengstrand_rotating_2019, gu_self-bound_2023}. Furthermore, assuming either a balanced mixture or a mean-field optimal ratio between particle numbers, the equations are solved for a single order parameter, so that vortices appear in both components, that is all vortices are empty. Under such conditions, various phases have been predicted~\cite{examilioti_ground_2020, nikolaou_rotating_2023, nikolaou_rotating_2023-1, tengstrand_rotating_2019, gu_self-bound_2023}, including vortices of single and multiple quantization and vortex lattices. 2D calculations predicted the stability of droplets with different vorticities in their components \cite{kartashov_2020}.

In the present work, we focus on potassium mixtures, using realistic interaction parameters for experimentally available magnetic fields aiming to determine the critical number of atoms needed to form a stable vortex and the nature of vortex states. Based on our previous work~\cite{sanuy_squeezing_2024}, we expect the critical number of atoms required to form a droplet with a vortex will reduce by squeezing the droplet in one direction. Following the procedure from Caldara and Ancilotto work \cite{caldara_vortices_2022}, we explore if empty or massive vortices represent the ground state and how this state evolves with the speed of rotation.
We study the droplets using the density functional theory. In Section 2 we define the system and explain the methods used to solve three dimensional extended Gross-Pitaevskii equations in rotating frame of reference. The results are presented in Section 3, while Section 4 gives the summary of results and conclusion.

\section{\label{sec:M}Method}

The droplets under consideration are binary mixtures of hyperfine states of ${}^{39}$K confined along the $z$-axis, which we consider within the framework of Density Functional Theory (DFT). The starting point is the energy functional,
 \begin{eqnarray}\label{eq:enfunc}
     E =&& \sum_i\int \dif\vec{r}\,\left[\frac{\hbar^2}{2m_i}|\nabla\psi_i(\vec{r})|^2 + \frac{1}{2}m_i \omega_z^2 z^2 \rho_i(\vec{r})\right]\;\;\;\nonumber\\
     &&+ \frac{1}{2}\sum_{i,j}g_{ij}\int\dif\vec{r} \rho_i(\vec{r})\rho_j(\vec{r}) \nonumber \\
     &&+ \int \dif\vec{r}\mathcal{E}_{\text{LHY}}[\rho_i(\vec{r}),\rho_j(\vec{r})],
 \end{eqnarray}
where $\rho_i(\vec{r}) = |\psi_i(\vec{r})|^2$ is the number density of component $i$ and $m_i$ is the mass. The interaction coupling constants are given by 
\begin{equation}
g_{ii} = \frac{4\pi\hbar^2 a_{ii}}{m_i}, \quad g_{12} = g_{21} = \frac{2\pi\hbar^2 a_{12}}{m_r},
\end{equation}
with $m_r$ being the reduced mass, and $a_{ij}$ the $s$-wave scattering lengths. Squeezing is achieved through a harmonic external potential with frequency $\omega_z = \frac{\hbar}{m a_z^2},$ where $a_z$ is the harmonic oscillator length. The oscillator length we use is $a_z = f \times 0.639\,\mu\text{m}$, where $f$ is the squeezing factor. The baseline oscillator length is taken from the  experiment \cite{cabrera_quantum_2018}, where they utilized an external potential with $a_{\text{ho}} = 0.639\,\mu\text{m}$. The same squeezing potential was used in the paper exploring the stability of squeezed droplets \cite{sanuy_squeezing_2024}.

The Lee-Huang-Yang (LHY) energy density is given by \cite{petrov_quantum_2015}
\begin{eqnarray} \label{eq:LHY-mix}
  &&  \mathcal{E}_{\text{LHY}}[\rho_1,\rho_2] =  \nonumber \\
   && \frac{8}{15\pi^2}\left(\frac{m_1}{\hbar^2}\right)^{\frac{3}{2}}(g_{11}\rho_1)^{\frac{5}{2}}
    F\left(\frac{m_2}{m_1}, \frac{g_{12}^2}{g_{11}g_{22}}, \frac{g_{22}\rho_2}{g_{11}\rho_1}\right),
\end{eqnarray}
where for the case of $m_1 = m_2$, the function $F(z = 1, u, x)$ is expressed as
\begin{equation}
    F(1,u,x) = \frac{1}{4\sqrt{2}}\sum_{\pm}\left[1 + x \pm \sqrt{\left(1 - x\right)^2 + 4ux}\right]^{\frac{5}{2}}.
\end{equation}
At the mean-field collapse, where $g_{12}^2 = g_{11}g_{22}$ or $u = 1$, this simplifies to
\begin{eqnarray}\label{eq:eLHY}
    &&\mathcal{E}_{\text{LHY}}[\rho_1,\rho_2] = \nonumber\\
    && \frac{8}{15\pi^2}\left(\frac{m_1}{\hbar^2}\right)^{\frac{3}{2}}(g_{11}\rho_1)^{\frac{5}{2}} 
    \left(1 + \frac{g_{22}\rho_2}{g_{11}\rho_1}\right)^{\frac{5}{2}}.
\end{eqnarray} 

Since we consider potassium mixtures of different hyperfine states, where $m_1 = m_2 = m$, we use Eq.~\eqref{eq:eLHY}. To the energy functional in Eq.~\eqref{eq:enfunc}, one can apply the variational principle, resulting in a coupled pair of generalized Gross-Pitaevskii equations (GPEs)
\begin{widetext}
\begin{equation} \label{eq:eGPE}
    i\hbar\frac{\partial}{\partial t}\psi_i (\vec{r},t) = \left[-\frac{\hbar^2}{2m}\nabla^2 + \frac{1}{2}m \omega_z^2 z^2 + g_{ii}|\psi_i(\vec{r},t)|^2 + g_{ij}|\psi_j(\vec{r},t)|^2 + \frac{\partial \mathcal{E}_{\text{LHY}}}{\partial \rho_i}\right]\psi_i (\vec{r},t) \equiv  \mathcal{H}_i\psi_i (\vec{r},t),
\end{equation}\end{widetext}
where $\mathcal{H}_i$ is the effective Hamiltonian for component $i$.

The total number of atoms is $N = N_1 + N_2$, with $N_i$ being the number of atoms of the $i$-th component. The optimal ratio of atom numbers~\cite{petrov_quantum_2015} , given by
\begin{equation}\label{eq:optimal}
\frac{N_1}{N_2} = \sqrt{\frac{a_{22}}{a_{11}}},
\end{equation}
 is generally used in our study. This ratio was confirmed as optimal in non-rotating vortex-free droplets by diffusion Monte Carlo (DMC) calculations \cite{cikojević2021abinitio}.

We explore droplets at three magnetic fields: 56.337 G, 56.453 G and 56.574 G; 
for which the scattering lengths are given in the Table~\ref{tab:scattering}. 
As a measure of the strength of the interaction, we define  $\delta a= 
a_{12}+\sqrt{a_{11}a_{22}}$.   When the field strength grows ($|\delta a|$ 
decreases), the droplet becomes more weakly interacting.

To find solutions for rotating droplets, Eq.~\eqref{eq:eGPE} is solved in the rotating frame of reference. This transformation leads to
\begin{equation}\label{eq:rotating_schrodinger_z}
    i\hbar \frac{\partial}{\partial t} \psi_i (\vec{r},t) = \left[\hat{\mathcal{H}}_i - \Omega \hat{L}_z \right] \psi_i (\vec{r},t),
\end{equation}
where $\Omega$ is the angular velocity and $\hat{L}_z$ is the $z$-component of 
the angular momentum operator. Solving Eq.~\eqref{eq:rotating_schrodinger_z} in 
imaginary time, $\tau = it$, allows for finding the ground state quite easily. 
In particular, we employ the split-step Fourier method method developed by Oktel 
\cite{oktel_vortex_2004}, and Chin and Krotscheck \cite{chin_fourth_order_2005}, 
which diagonalizes part of the Hamiltonian and thus allows for a fast and 
efficient time evolution. Since our droplets are not confined in the $xy$-plane, 
we add and subtract a weak harmonic potential to facilitate the canonical 
transformation. We use $a_{xy}=1000\,\mathrm{\mu}$m, in order to minimize the 
impact on the density profile and energy, and anisotropy $\delta = 0$. We 
use the time step $dt = 100 \frac{ma_{11}^2}{\hbar}$ and a grid size $256 \times 
256 \times 64$. To reduce computation time, for vortex-free droplets, we use the 
smaller grid size $64 \times 64 \times 64$ to get a coarse ground state which we 
interpolate to the finer grid size. Finally, we propagate that for a short time 
to smooth it over. Further details about the method may be found in Appendix 
\ref{app:numerical}.
\begin{table}[t]
    \caption{\label{tab:scattering} Scattering lengths $a_{ij}$ and the $\delta a$ value for the $^{39}$K mixture (in units of Bohr's radius $a_0$) as a function of magnetic field $B$ (in G).}
        \begin{ruledtabular}
            \begin{tabular}{ccccc}
                $B\,(\text{G})$ & $a_{11}\,(a_0)$ & $a_{22}\,(a_0)$ & $a_{12}\,(a_0)$ & $\delta a\,(a_0)$ \\\hline
            56.337 & 66.619 & 34.369 & -53.386 & -5.536 \\
            56.453 & 70.119 & 34.136 & -53.333 & -4.409 \\
            56.574 & 74.118 & 33.895 & -53.278 & -3.156 \\
        \end{tabular}
        \end{ruledtabular}
\end{table}

As an initial condition for vortex-free droplets, we use a Gaussian which quickly converges to the ground state, whose energy and density profile agree with previous calculations~\cite{sanuy_squeezing_2024}. To find states with vortices, we take the vortex-free solution and introduce a vortex in the center as
\begin{equation} \label{eq:wfvortex}
    \psi_j = \psi_{j,\text{no vortex}} \times e^{il_j\phi},
\end{equation}
where $\phi = \arctan\left(\frac{y}{x}\right)$ is the polar angle and $l_j$ being the integer quantum number of the vortex in component $j$, here taken to be 0 (vortex absent) or 1 (vortex present). Vortex states appear highly stable even when they are excited states. This permits calculations of energies of states with a vortex present in one, the other, or both components.

\section{\label{sec:R}Results}
\begin{figure*}[t]
    \centering
    \includegraphics[width=\linewidth]{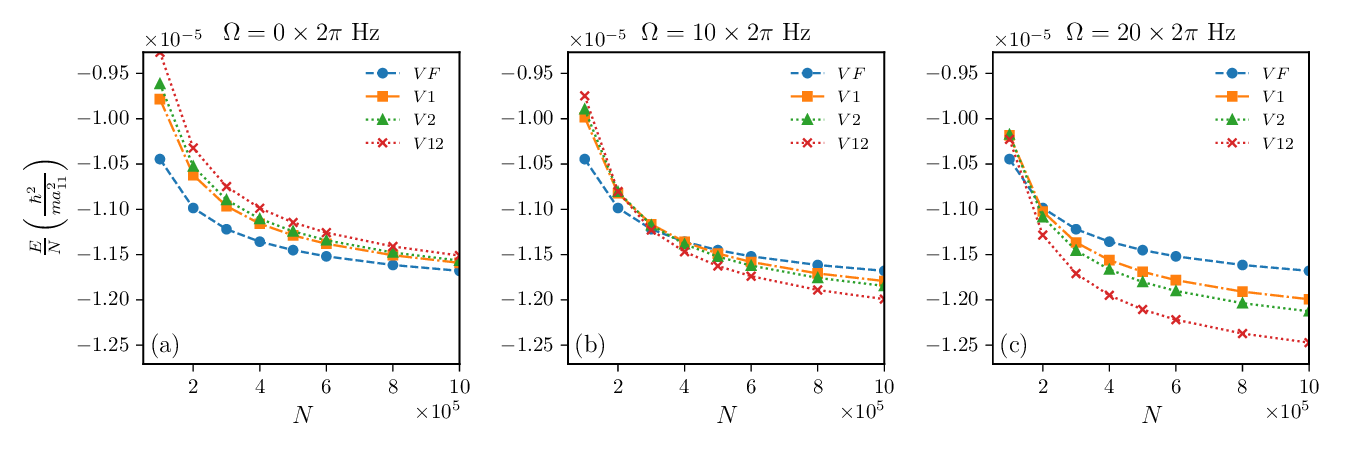}
    \caption{\label{fig:E/337_025} Energy per atom as a function of the number of atoms $N$ for droplets at magnetic field $B = 56.337$ G ($\delta a = -5.536\, a_0$) at squeezing $f=0.25$, plotted for (a) non-rotating droplets and droplets rotating at (b) $10\times 2\pi$ Hz and (c) $20\times 2\pi$ Hz. $VF$ denotes a vortex-free droplet, while $V1$ and $V2$ denote vortices in the first and second components, respectively, and $V12$ denotes 2 vortices, one in each component.}
\end{figure*}
The condition for stability of Bose-Bose quantum droplets is the requirement of having more atoms than a certain critical value, $N_c$. In the Introduction, we noted that for squeezed \({}^{39}\)K droplets the critical number of atoms had been previously determined through DFT calculations \cite{sanuy_squeezing_2024} to be on the order of $\sim 10^3$.
Droplets with $N>N_c$ are \emph{self-bound} systems, meaning the energy $ E = E_\text{total} - E_\text{HO} $ is negative. Here \(E_\text{HO}\) is the single-particle energy contribution of the squeezing potential. The $N>N_c$ condition holds in all cases we examined in the present work. For illustration, we present in Fig.~\ref{fig:E/337_025} the energy per particle for droplets at a magnetic field of \( B = 56.337\,\mathrm{G} \) and a squeezing factor \( f = 0.25 \) in the non-rotating case and in the case of rotation with angular velocities $10\times 2\pi$ Hz and $20\times 2\pi$ Hz. In all shown cases, the droplets remain self-bound.

Analyzing the energies for rotating systems shown in Fig.~\ref{fig:E/337_025}, one can notice crossover points in the total number of atoms \( N \), beyond which the \( V12 \) state, characterized by two overlapping vortices, one in each component, becomes the energetically favored ground state. This transition occurs around \( N = 285000 \) and \( N = 132000 \) atoms, for angular velocity $10\times 2\pi$ Hz and $20\times 2\pi$ Hz, respectively, marking the critical size for vortex hosting. We conclude, in line with expectations, that faster rotation eases vortex hosting.

On the other hand, hosting a single vortex requires significantly larger number of particles than those for having self-bound droplets. For heteronuclear mixtures of ${}^{41}\text{K}$-${}^{87}\text{Rb}$, this critical number has been estimated to be of the order of $10^6$ particles \cite{caldara_vortices_2022}. The presence of vortices increases the critical size due to the additional kinetic energy and internal structure.

\begin{figure}[t]
    \centering
    \includegraphics[width=\linewidth]{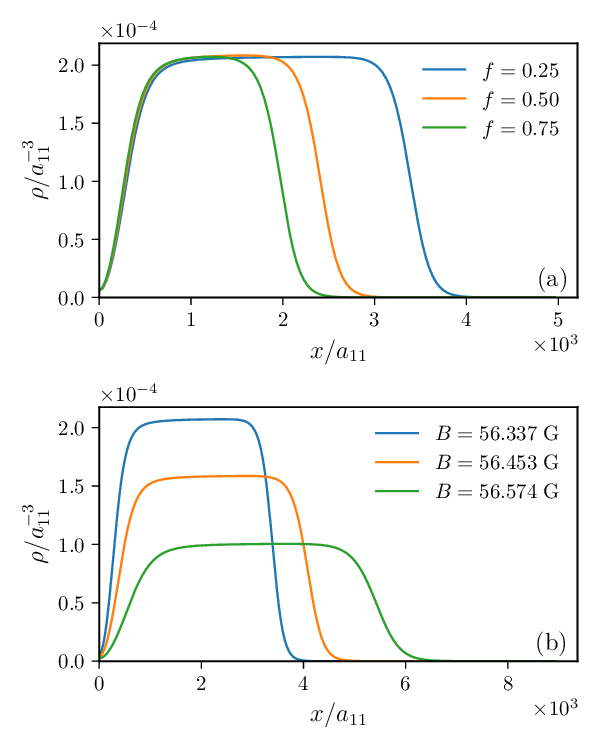}
    \caption{\label{fig:den/600k_V2_squeezing} Total density $\rho(x,0,0)$ profiles along the line passing through the center for a droplet with a vortex in component 2 ($V2$). Droplets are made up from $N=600000$ atoms, and are rotating with angular velocity $\Omega = 20 \times 2\pi$ Hz. (a) Plotted at magnetic field $B = 56.337$ G and three different squeezings,  $f=$ 0.25, 0.50, and 0.75.  (b) Plotted at different magnetic fields, at squeezing $f=0.25$. All $a_{11}$ units correspond to the field $B=56.337$ ($\delta a = -5.536\, a_0$).}
\end{figure}

More squeezed droplets can facilitate the formation of vortices. To explain this feature, we consider density profiles. In Ref.~\cite{sanuy_squeezing_2024}, it was shown that squeezed droplets reach their bulk density sooner, and consequently their spatial extent perpendicular to the squeezing direction is larger. This larger radius makes it easier to accommodate the vortex core.
In Fig.~\ref{fig:den/600k_V2_squeezing}a we show the density profile for a droplet hosting a vortex in the more numerous component 2, along the axis perpendicular to the squeezing, at three different squeezing strengths. Notably, regardless of the confinement along the $z$-axis, the droplets always reach the same bulk density. This leads to more strongly confined droplets having a larger radius in the perpendicular $xy$-plane. The same figure, ~\ref{fig:den/600k_V2_squeezing}b, also shows the droplet of the same size at the strongest squeezing considered for three different magnetic fields. We observe that for weaker interaction (stronger $B$) the central densities are smaller and droplets are more extended. Also, the vortex cores decrease with increasing interaction, which was also observed in $^{41}$K-$^{87}$Rb mixtures~\cite{caldara_vortices_2022}.

\begin{figure}[t]
    \centering
    \includegraphics[width=\linewidth]{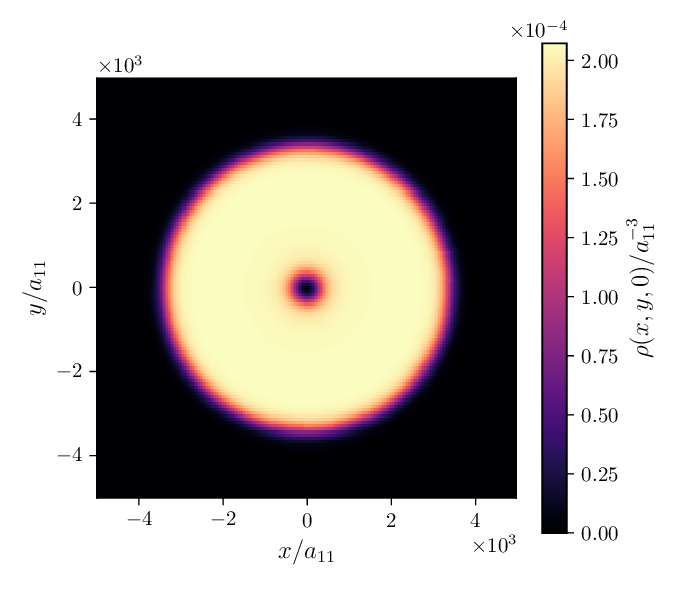}
    \caption{\label{fig:den/600k_V2_0.25_2d} 2D total density profile in the plane passing through the center for a droplet with a vortex in component 2. The droplet is made up from $N=600000$ atoms at magnetic field $B = 56.337$ G ($\delta a = -5.536\, a_0$) and squeezing $f=0.25$. The droplet is rotating with angular velocity $\Omega = 20 \times 2\pi$ Hz.}
\end{figure}

\begin{figure}[t]
    \centering
    \includegraphics[width=\linewidth]{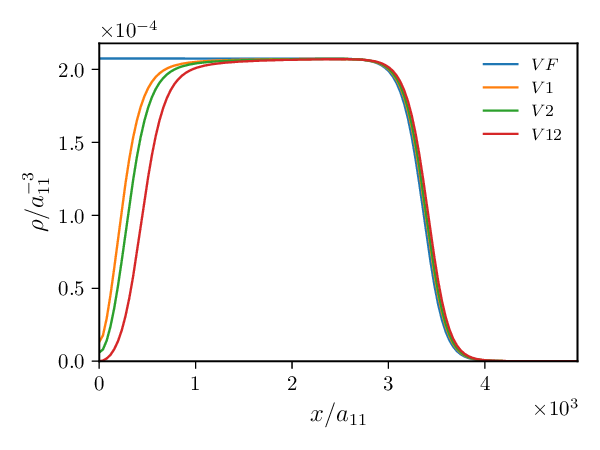}
    \caption{\label{fig:den/600k_0.25_V} Total density profiles along the line passing through the center for a droplet made up from $N=600000$ atoms at magnetic field $B = 56.337$ G ($\delta a = -5.536\, a_0$) and squeezing $f=0.25$. The droplet is rotating with angular velocity $\Omega = 20 \times 2\pi$ Hz. $VF$ denotes a vortex-free droplet, while $V1$ and $V2$ denote vortices in the centers of the first and second components, respectively, and $V12$ denotes 2 vortices, one in each component.}
\end{figure}

An example of the 2D density profile in the $xy$ plane is shown in 
Fig.~\ref{fig:den/600k_V2_0.25_2d}. The droplet hosts a central vortex in the 
second component, and its density in the center reaches a finite value. This is 
not the case when two overlapping vortices, one in either species, are present, 
as exhibited in Fig.~\ref{fig:den/600k_0.25_V}. The density of the component 
carrying vorticity vanishes in the vortex core, while the other component's 
density reaches a finite value. That is why in Fig.~\ref{fig:den/600k_0.25_V}, 
in cases $V1$ and $V2$, we see finite values reached, while for $V12$ it goes to 
zero. Concretely, in the $V2$ case, the density of component 1 reaches 6-7 \% of 
the bulk density value depending on the magnetic field. This value is lower than 
the reported central density of $^{41}$K-$^{87}$Rb in  droplets (around 18\%)  
\cite{caldara_vortices_2022}, which could be due to a more negative $\delta a$ 
than in our work, as Ref. \onlinecite{caldara_vortices_2022} states that the 
amount of filling reduces with $a_{12}$ becoming less negative. The observed 
filling of vortices is, however, significantly lower than in repulsive mixtures, 
where it is even possible for the component filling the vortex to achieve 
maximum density within the vortex, thus forming the so-called coreless 
vortices~\cite{Kasamatsu_2005,christensson_2008}.

\begin{figure}[t]
    \centering
    \includegraphics[width=\linewidth]{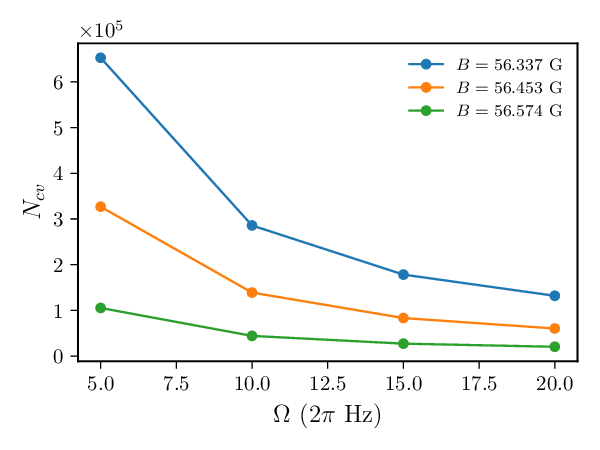}
    \caption{\label{fig:Nc/025} Critical number of atoms for hosting a vortex, $N_{cv}$, as a function of angular velocity $\Omega$, for a droplet at three magnetic fields $B = 56.337$ G, $B = 56.453$ G and $B = 56.574$ G, at squeezing $f=0.25$.}
\end{figure}

As noted earlier, at stronger fields atoms interact more weakly (smaller $|\delta a|$). This  is reflected in smaller self-binding energies and larger critical atom number, $N_c$, for forming a self-bound droplet.~\cite{sanuy_squeezing_2024} Interestingly, the weaker interparticle interactions require lower number of atoms needed for hosting a vortex, $N_{cv}$, as shown in Fig.~\ref{fig:Nc/025}. For the strongest squeezing and field considered, and for the largest angular velocity, the critical atom number drops to around 25000 atoms.
To explain the reduction in the critical atom number for vortex hosting, it is important to note that when vortices are stable the central densities of drops approaches the bulk density, which is significantly smaller at stronger fields, as shown in Fig.~\ref{fig:den/600k_V2_squeezing}. At the same time |$E/N$| is smaller, as well as the difference of energies per particle with and without vortex not considering the rotational contribution, that is $E'_{Vi}/N-E'_{VF}/N$, with $E' = E - E_\text{rot}$. The rotational energy per particle in droplets without vortex is essentially zero, while for droplets with vortex it is the same for each $B$,  $-\Omega L_z/N=-\Omega \hbar$, which means it manages to lower the energy of the state with vortex below the state without vortex sooner (for smaller $N$) in droplets with smaller $|E/N|$, that is, for droplets in stronger fields. Therefore, despite the larger cores more weakly interacting droplets are more extended, as shown in Fig.~\ref{fig:den/600k_V2_squeezing} and reach conditions for having a stable vortex sooner.

\begin{figure}[]
    \centering
    \includegraphics[width=\linewidth]{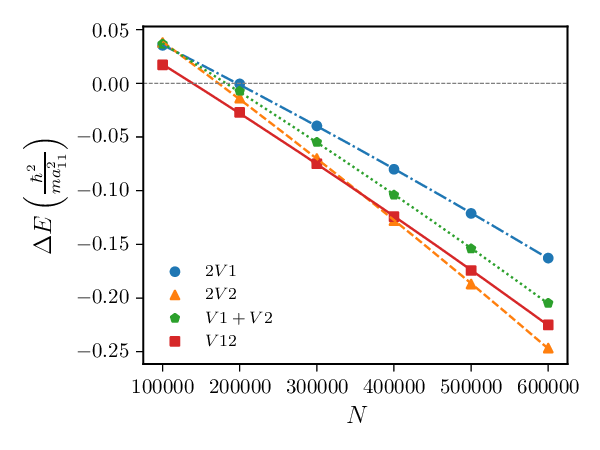}
    \caption{\label{fig:2DE/543_025_om_10} Energy difference between the vortex states and vortex-free states $\Delta E$, as a function of the number of atoms $N$ in a droplet at magnetic field $B = 56.453$ G ($\delta a = -4.409\, a_0$) and squeezing $f=0.25$, rotating with angular velocity $\Omega = 10 \times 2\pi$ Hz. The depicted fit is second order polynomial. $2V1$ and $2V2$ denote two separated vortices in components 1 and 2. $V12$ corresponds to two overlapping central vortices, one in each component, and $V1+V2$ two separated vortices which do not overlap, also one in each component.}
\end{figure}

In aforementioned droplets we have compared energies per particle when hosting a single vortex in component 1 or component 2, and having a vortex in both. In order to have the same vorticity in all cases, it is more apt to compare energies needed to form two vortices in components 1 or 2 with a case of a vortex in both. To that end we assume that vortices are not interacting, and in a similar vein to Ref.\cite{caldara_vortices_2022} we compare the following energies
\begin{eqnarray}
    \left.\Delta E\right|_{2Vi} &=& 2 (E_{Vi} - E_{VF})\,,\\
    \left.\Delta E\right|_{V12} &=& E_{V12} - E_{VF}\,, \\
    \left.\Delta E\right|_{V1 + V2} &=& E_{V1} + E_{V2} - 2E_{VF}\,.
\end{eqnarray}
Here, $2Vi$, $i=1,2$ signifies two separated vortices in component $i$. $V12$ corresponds to two central overlapping vortices, one in each component. Finally, $V1+V2$ marks two vortices, one in each component, which do not overlap. By comparing these energies, we aim to gain insight into the ground state of the droplets.
This comparison relies upon several assumptions. Primarily, it assumes that two non-overlapping vortices can fit into our droplet, which becomes less of a concern as \(N\) increases, and secondly, that the vortices are not interacting.
We plot the said energies in Fig.~\ref{fig:2DE/543_025_om_10} for droplets at the magnetic field strength of \( B = 56.453\,\mathrm{G} \) with the squeezing factor of \( f = 0.25 \), rotating with  \( \Omega = 10 \times 2\pi\,\mathrm{Hz} \). The first observation is that at low atom numbers, none of the energy differences is negative, meaning that the vortex-free ($VF$) state is the ground state. Then, $V12$ becomes the ground state at around \(N = 140 000\), and persists being so until $2V2$ state overtakes it at around \(N = 360 000\). 

\begin{figure*}[t]
    \centering
    \includegraphics[width=\linewidth]{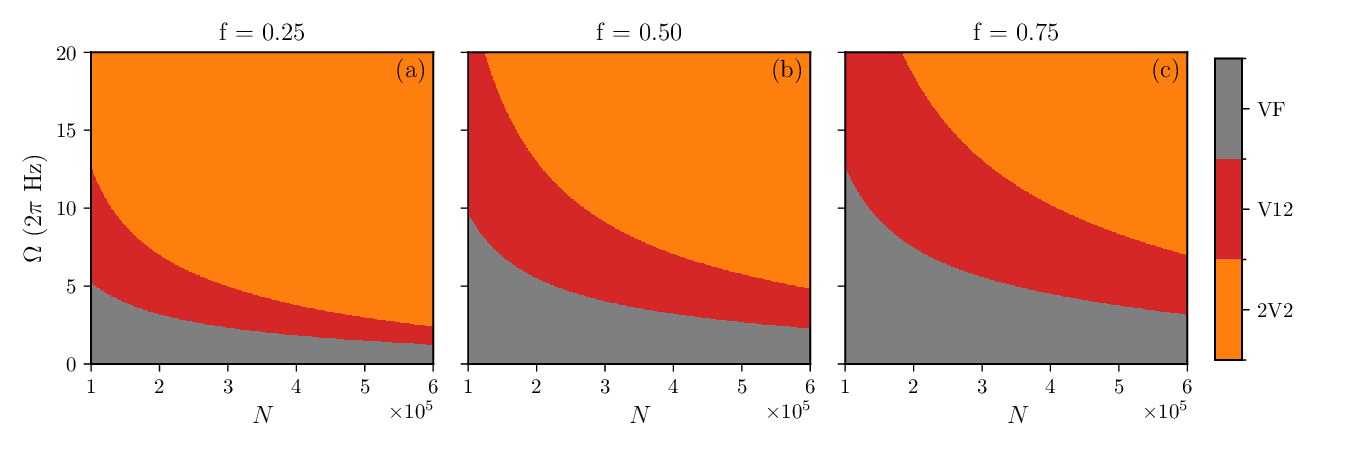}
    \caption{\label{fig:Diag_N_om_574} Diagrams showing the energetically most favorable state for a droplet at $B = 56.574$ G ($\delta a = -3.156\, a_0$), as it changes with number of atoms $N$ and angular velocity $\Omega$, for three different squeezings, $f=$ (a) 0.25, (b) 0.50, and (c) 0.75.}
\end{figure*}

We can fit the data obtained for $\Omega =$ 0, 5, 10, 15 and 20 $\times 2\pi$ Hz, to second order polynomials, and do the same comparison for a grid of $N$ and $\Omega$ values. The result are diagrams, shown in Fig.~\ref{fig:Diag_N_om_574} for droplets at the field $B=56.574$ G, for different confinements. At slow rotation or with a small number of atoms, the vortex-free state is the ground state. By increasing $\Omega$ for a fixed $N$, or by adding more atoms at a fixed $\Omega$, the state $V12$, one with an empty vortex, becomes the ground state. That is, vorticity will be distributed across both components. Further increases in $N$ or $\Omega$ lead to states where vortices in the second, more numerous species become favorable.
We also illustrate the effect of squeezing: the stronger the confinement, the less atoms are needed to host a vortex.
From our data we have observed that with weaker interactions it becomes possible to host vortices with fewer atoms or with slower rotation. At larger $N$, the $V12$ zone is noticeably smaller. 
To check the sensitivity of the diagrams to the assumption of non-interacting vortices, we performed several imaginary time calculations with $2V_i$ configurations and found the diagrams to be essentially unchanged, with only a small increase of the stability range of the $V12$ configurations.


Instead of having two vortices of single vorticity ($l$=1), one can also have a single vortex with $l=2$. In homogeneous superfluids multiply quantized vortices are unstable because the kinetic energy grows with $l^2$, while the rotational energy grows with $l$. Putting $l_j=2$ in Eq. \ref{eq:wfvortex} we have calculated the energies of droplets with squeezing 0.25. For large enough droplets, the energies of the droplet with $l$=2 vortex at the origin in component $i$   $E_{Vi}^{l=2}$ becomes lower than the energy of the single vortex with $l$=1 $E_{Vi}^{l=2}<E_{Vi}$, but $E_{Vi}^{l=2}-E_{VF} > \left.\Delta E\right|_{2Vi}$, which means it is more energetically favorable to have two vortices in component $i$ than one with $l=2$.   In all cases, we found the droplets hosting multivortex to be metastable, splitting after very long imaginary time simulation into vortices with $l=1$.  

Experimentally, it is challenging to maintain the optimal population balance. 
To investigate 
the effects of the population imbalance, we took \(N_2/N_1 = x 
\sqrt{a_{11}/a_{22}} \), with \(x=0.8\) and \(x=1.2\). The choice was motivated 
by the estimates of probable population imbalance in Taruell's 
experiment~\cite{cabrera_quantum_2018} made in 
Ref.\onlinecite{cikojevic2020finite}. Test calculations were performed with  
\(\Omega = 10\times 2\pi\) Hz, \(f = 0.5\), \(B = 56.574\) G, for \(N = 100000,\; 
300000,\; 600000\) droplets. We found that the droplets were less bound with 
population imbalance, as expected. The energies increased from 12 to 26\%. 
However, the qualitative behavior of the droplets was unchanged. That is, the 
$V12$ state became favorable first and was subsequently overtaken by $2V2$ 
around the same droplet size for all population ratios considered (the droplet 
size at crossover from $V12$ to $2V2$ is shifted by about +2\% for $x=0.8$ and 
by about -8\% for $x=1.2$ with respect to $x=1$). Therefore, only small 
quantitative changes are expected in the phase diagrams presented in Fig. 
\ref{fig:Diag_N_om_574}.

\begin{figure*}[t]
    \centering
    \includegraphics[width=\linewidth]{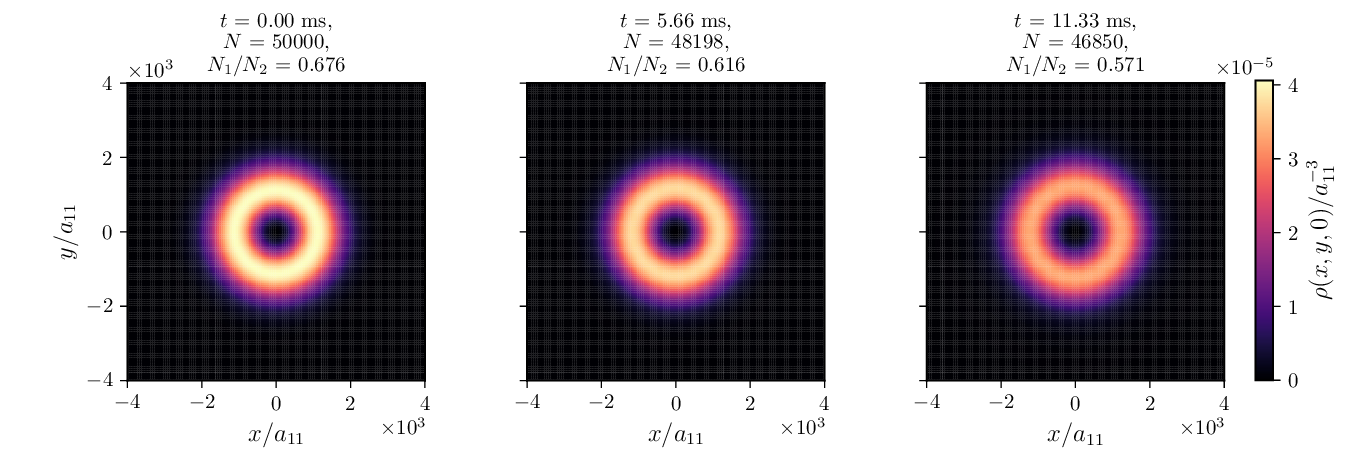}
    \caption{\label{fig:2d_den_time_V12} 2D total density profiles in the plane passing through the center for a droplet with a vortex in both components ($V12$). The droplet is initially composed of $N=50000$ atoms, and is propagated in real time with three body losses present. Magnetic field used is $B = 56.574$ G ($\delta a = -3.156\, a_0$) with squeezing $f=0.5$, rotating with angular velocity $\Omega = 30 \times 2\pi$ Hz.}
\end{figure*}

Another, very important experimental challenge is the finite lifetime of 
droplets, 
on the order of ten milliseconds, as reported in Ref. 
\onlinecite{cabrera_quantum_2018}. It is caused by three-body recombination, 
which primarily affects the first component. To verify whether this is a 
limiting factor for observing the vortices, we performed some real-time 
evolutions of droplets, with included three-body losses for the first 
component, similarly to what was made 
in Ref.~\onlinecite{cikojevic2021collision}. This was achieved by adding a term 
$-\frac{i}{2}\hbar\kappa\rho_1^2$ to the potential part of the Hamiltonian, with 
\(\kappa\) being the three-body loss coefficient. Since we are considering 
squeezed droplets, we used the value quoted in Ref. 
\onlinecite{cabrera_quantum_2018}, $\kappa = 7.5 \times 
10^{-28}\text{cm}^6/\text{s}$, which is reported to be uncertain up to the 
factor of 2. We investigated several test cases of droplets, with and without 
vortices, and found the droplets with vortices to have more than double 
lifetime of the same initial size droplets without vortices. This is most likely 
due to the lower density in the vortex core, which leads to smaller losses 
($\propto\rho_1^2$). The state $V12$ lasts longer than the state $V2$, due to 
lower core density $\rho_1$ when compared to the $V2$ or $VF$ case. In Figure 
\ref{fig:2d_den_time_V12} we report a real-time evolution of a droplet with 
initially 50000 atoms, in a magnetic field $B$=56.574 G, $\Omega = 30\times 
2\pi$ Hz and squeezing $f$=0.5, with a vortex in both components. It starts to 
lose atoms mainly in the $N_1$ component, slowly increasing its energy and 
decreasing $N_1/N_2$. At the end of the simulation time of 1.5$\times$10$^6$, 
which corresponds to $14.16$ ms, the droplet was still self-bound, while the 
same size droplet without the vortex evaporated at 
$6.61$ ms.  Typically, droplets stopped being self-bound at  $N_1/N_2 \approx 
0.5$.   The lifetime of the droplets is increased for higher angular velocity, 
as it takes longer for the self-binding energy to reach zero.  Squeezing the 
droplets of a given number of atoms more strongly shortens their lifetime, most 
likely due to larger part of the drop being saturated. Also, droplets with 
weaker interactions (smaller $|\delta a|$) have smaller central densities and 
thus longer lifetimes. 

\section{\label{sec:C}Conclusions}

In this study, we examined the energies of vortex-hosting weakly bound squeezed Bose-Bose potassium droplets using DFT simulations and realistic interaction parameters for experimentally accessible magnetic fields. The primary objective was to investigate the conditions under which vortex states become the ground states, particularly at various angular velocities and confinement strengths.

We determined the critical number of atoms necessary for these vortex states to emerge as ground states, and found it to be on the order $\approx 10^4-10^5$, noting that it is lower in droplets that rotate faster and in those with weaker interactions, and that it reduces when squeezing the droplet in one direction. Due to using a three-dimensional density functional we have not explored further squeezing towards two-dimensional limit, but it is expected that it would further reduce the critical atom number for hosting a vortex $N_{cv}$. The predicted values for $N_{cv}$ are within the experimental reach and can help guide the experimental efforts.

Under the assumption that vortices do not interact, we identified a distinct region where empty vortices, with a central vortex present in both components, are the ground state. This is in contrast to the more commonly observed filled (massive) vortices. Empty vortices are the stable ground state in smaller and more slowly spun droplets. With the increase in droplet size and angular velocity it becomes energetically favorable for the droplet to host vortices in more numerous component, which results in filled vortices.

We have not found conclusive evidence for stability of vortices with multiple vorticity, which are predicted for repulsive trapped Bose-Bose mixtures~\cite{Bargi_2007,christensson_2008, Kuopanportti_2015, Patrick_2023} as a consequence of interspecies interaction. One of the reasons could be that the filling of vortices is much weaker than in the case of repulsive mixture. 

Based on our results in selected cases, we do not expect that our predictions 
would  change significantly in the case of population imbalance. Importantly, 
we showed that droplets with vortices have longer lifetimes compared 
to vortex-free droplets, making their detection experimentally feasible despite 
three-body losses.

\begin{acknowledgments}
We thank Francesco Ancilotto for insightful discussion.
We acknowledge financial support from the Croatian Science Foundation project IP-2022-10-6144 
and from Ministerio de Ciencia e
Innovaci\'on MCIN/AEI/10.13039/501100011033
(Spain) under Grant No. PID2023-147469NB-C21  and
from AGAUR-Generalitat de Catalunya Grant No. 2021-SGR-01411.
	The computational resources of UniST-Phy server at the University of Split and supercomputer Supek at SRCE in Zagreb were used.

\end{acknowledgments}

\section*{DATA AVAILABILITY}
The data that support the findings of this article are openly available.~\cite{Poparic}

\appendix
\section{\label{app:numerical}Details of the numerical method}

The Hamiltonian in the rotating frame, given by Eq.~\eqref{eq:eGPE}, takes the form
\begin{eqnarray}
    \label{eq:H.sys}
    \mathcal{H} = &&\frac{1}{2m}\left(p_x^2 + p_y^2 + p_z^2\right) \nonumber\\
    &&+ \frac{1}{2}m\omega_{z}^2z^2 + V_{\text{MF+LHY}} - \Omega\left(xp_y - yp_x\right),
\end{eqnarray}
where $V_{\text{MF+LHY}}$ is the combined mean-field and Lee-Huang-Yang potential.

To solve this Hamiltonian, we add and subtract a potential in the $xy$-plane and split it as follows
\begin{eqnarray}\label{eq:H.split}
    \mathcal{H} &=& \mathcal{H}_{xy} + \mathcal{V} + \mathcal{T}_z, \\
    \mathcal{H}_{xy} &=& \frac{1}{2m}\left(p_x^2 + p_y^2\right) \nonumber\\ 
    &&+ \frac{1}{2}m\omega_{xy}^2\left[(1+\delta)x^2 + (1-\delta)y^2\right] \nonumber\\
    &&- \Omega\left(xp_y - yp_x\right), \\
    \mathcal{V} &=& - \frac{1}{2}m\omega_{xy}^2\left[(1+\delta)x^2 + (1-\delta)y^2\right] \nonumber \\
    &&+ \frac{1}{2}m\omega_{z}^2z^2 + V_{\text{MF+LHY}}, \\
    \mathcal{T}_z &=& \frac{1}{2m}p_z^2.
\end{eqnarray}
We use units of $\hbar^2/(ma_{11}^2)$ for the Hamiltonian and energy. Consequently, the coordinates and momenta are naturally redefined as $x' = x/a_{11}$ and $p_x' = -i a_{11}\frac{\partial}{\partial x}$,
where $a_{11}$ is the scattering length between the atoms of the first component. In reciprocal space, the momentum simplifies to $p_x' = k_x'$. For simplicity, we drop the primes in subsequent equations. We also express the $xy$ potential using the harmonic oscillator length $a_{xy}$, defined by the relation $\omega_{xy} = \frac{\hbar}{m a_{xy}^2}$.

To solve the part of the equation associated with $\mathcal{H}_{xy}$, we employ the method of Oktel \cite{oktel_vortex_2004}, and Chin and Krotscheck \cite{chin_fourth_order_2005}. Using the following linear canonical transformation:
\begin{eqnarray}
    Q_1 &=& \alpha_1 \left[\cos(\phi)\tilde{x} - \sin(\phi)p_y\right], \\
    P_1 &=& \frac{1}{\alpha_1} \left[\sin(\phi)\tilde{y} + \cos(\phi)p_x\right], \\
    Q_2 &=& \alpha_2 \left[\cos(\phi)\tilde{y} - \sin(\phi)p_x\right], \\
    P_2 &=& \frac{1}{\alpha_2} \left[\sin(\phi)\tilde{x} + \cos(\phi)p_y\right],
\end{eqnarray}
with the condition that $\tan(2\phi) = 2\tilde{\Omega}/\delta$, we can diagonalize $\mathcal{H}_{xy}$. Here, $\tilde{x} = x/a_{xy}^2$ and $\tilde{\Omega} = f_{\Omega}\Omega = \frac{ma_{11}^2}{\hbar}\Omega$.

Applying this transformation, the Hamiltonian $\mathcal{H}_{xy}$ takes the diagonal form:
\begin{equation}\label{eq.H.xy.diag}
    \mathcal{H}_{xy} = \frac{1}{2}\left[P_1^2 + P_2^2 + \Omega_1^2Q_1^2 + \Omega_2^2Q_2^2\right],
\end{equation}
where the coefficients are given by:
\begin{eqnarray}
    \frac{1}{\alpha_1^2} &=& 1 - \frac{\delta}{2} + \frac{1}{2}\sqrt{\delta^2 + 4\tilde{\Omega}^2}, \\
    \frac{1}{\alpha_2^2} &=& 1 + \frac{\delta}{2} - \frac{1}{2}\sqrt{\delta^2 + 4\tilde{\Omega}^2}, \\
    \Omega_1^2 &=& \frac{1}{\alpha_1^2} \left(1 + \frac{\delta}{2} + \frac{1}{2}\sqrt{\delta^2 + 4\tilde{\Omega}^2}\right), \\
    \Omega_2^2 &=& \frac{1}{\alpha_2^2} \left(1 - \frac{\delta}{2} - \frac{1}{2}\sqrt{\delta^2 + 4\tilde{\Omega}^2}\right).
\end{eqnarray}

We use a second order algorithm in imaginary time $\tau=it$ from \cite{chin_fourth_order_2005}
\begin{eqnarray}
		\psi(\tau+\Delta\tau) =&&e^{-\Delta\tau\mathcal{T}_z}
        e^{-\frac{1}{6}\Delta\tau\mathcal{V}}e^{-\frac{1}{2}\Delta\tau\mathcal{H}_{xy}} \nonumber \\
        &&e^{-\frac{2}{3}\Delta\tau\mathcal{V}}e^{-\frac{1}{2}\Delta\tau\mathcal{H}_{xy}}
        e^{-\frac{1}{6}\Delta\tau\mathcal{V}} \psi(\tau).
\end{eqnarray}
Evolution with $\mathcal{H}_{xy}$ is done in steps, as detailed in \cite{chin_fourth_order_2005}:\begin{enumerate}
	\item Starting from $\psi(x,y,z)$ we compute $\psi(p_x,y,z)$ using a one-dimensional fast Fourier transform (FFT) and multiply the result by $\exp\left[-\frac{\Delta\tau}{8}\left(P_1^2 + \Omega_2^2Q_2^2\right)\right]$.
	\item We compute $\psi(x,p_y,z)$ using a two-dimensional FFT and multiply the result by$\exp\left[-\frac{\Delta\tau}{4}\left(\Omega_1^2Q_1^2 + P_2^2\right)\right]$.
	\item We compute $\psi(p_x,y,z)$ using an inverse two-dimensional FFT and multiply the result by $\exp\left[-\frac{\Delta\tau}{8}\left(P_1^2 + \Omega_2^2Q_2^2\right)\right]$.
	\item We compute $\psi(x,y,z)$ using the one-dimensional inverse FFT.
\end{enumerate}
The algorithm remains stable for $\Omega < 1/f_\Omega$ when $\delta = 0$ and taking $\Omega = 0$ reduces the problem to a non-rotating one, with $Q_1 = \tilde{x}$, $P_1 = p_x$, $Q_2 = \tilde{y}$ and $P_2 = p_y$

\end{document}